\documentclass[preprint,prd,nofootinbib,tightenlines,amsmath]{revtex4-1}
\usepackage{bbm}
\usepackage{amsfonts}
\usepackage{mathrsfs}
\usepackage{graphicx}
\usepackage{bm}
\usepackage{hyperref}    

\begin{document}
\baselineskip=15pt \parskip=5pt

\vspace*{3em}

\title{Interpretation of the gamma-ray excess and AMS-02 antiprotons: Velocity dependent dark matter annihilations}

\author{Lian-Bao Jia}
\email{jialb@mail.nankai.edu.cn}

\affiliation{School of Science, Southwest University of Science and Technology, Mianyang
621010,  China \\
 }

\begin{abstract}

The two messenger results of the GeV gamma-ray excess at the Galactic center and a probable antiproton excess in the recent AMS-02 observation suggest that these two anomalies may be owing to the same origin --- the dark matter (DM) annihilation into $b \bar b$, while these results seem in tension with the dwarf spheroidal galaxy observations. To give a compatible explanation about it, we consider the pseudoscalar DM particles $S_d^+ S_d^-$ annihilating via $S_d^+ S_d^-  \rightarrow S_d^0 S_d^0$, with the process mediated by a new scalar $\phi$ and $S_d^0$ quickly decaying into $b \bar{b}$. For the particles $S_d^+$, $S_d^-$ and $S_d^0$ in a triplet with degenerate masses, the annihilation cross section of DM today is linearly dependent on the relative velocity $v_r$, and thus constraints from the dwarf spheroidal galaxies are relaxed. The parameter spaces are derived with corresponding constraints. Though traces from the new sector seem challenging to be disclosed at collider and in DM direct detections, the indirect search of the gamma-ray line from the $S_d^0$'s decay has the potential to shed light on DM annihilations, with the energy of the gamma-ray line $\sim m_{S_d^0} /2$, i.e. about 50$-$75 GeV.

\end{abstract}

\maketitle

\section{Introduction}

Today the nature of dark matter (DM) is still unclear, and the cosmic ray observation may indirectly provide some properties of DM. One possible DM signature is the Galactic center (GC) 1$-$3 GeV gamma-ray excess, which may be due to the weakly interacting massive particle (WIMP) type DM annihilating into $b \bar b$, $c \bar c$, $q \bar q$ (light quarks $u$, $d$, $s$), $\tau \bar{\tau}$ et al. \cite{Goodenough:2009gk,Hooper:2010mq,Hooper:2011ti,Abazajian:2012pn,Hooper:2012sr,Abazajian:2014fta,Daylan:2014rsa,Alves:2014yha,Zhou:2014lva,Calore:2014xka,Agrawal:2014oha,Calore:2014nla,TheFermi-LAT:2015kwa,Huang:2015rlu,Karwin:2016tsw}, e.g. WIMPs in a mass range about 35$-$74 GeV annihilating into $b \bar b$ with the cross section about (1$-$3)$\times 10^{-26}$ cm$^3$/s. Meanwhile, the updated constraints of the dwarf spheroidal galaxies from the Fermi-LAT \cite{Ackermann:2015zua,Li:2015kag,Fermi-LAT:2016uux} seem in tension with most DM interpretations of the GC gamma-ray excess. Yet, some schemes can also be compatible with the constraints of dwarf galaxies, such as WIMPs annihilating into $\mu^+ \mu^-$, $e^+ e^-$ \cite{Liu:2014cma,Kaplinghat:2015gha}, decays of the asymmetric DM with anti-DM holding enough energy to escape dwarf spheroidal galaxies \cite{Hardy:2014dea}, or the p-wave annihilating DM from a decaying predecessor \cite{Choquette:2016xsw}. Another possible explanation about the gamma-ray excess is the millisecond pulsars \cite{Hooper:2010mq,Hooper:2011ti,Abazajian:2012pn,Abazajian:2010zy,Wharton:2011dv,O'Leary:2015gfa,Bartels:2015aea,Lee:2015fea}, while some discussions \cite{Cholis:2014lta,Linden:2015qha,Hooper:2016rap,Haggard:2017lyq} indicate that this astrophysical explanation seems challenging to produce the majority of the observed gamma-ray excess.

In addition, the charged cosmic rays may also shed light on the properties of DM. Recently, the analysis of the antiproton flux from the AMS-02 observations \cite{Aguilar:2016kjl} indicates the existence of a possible DM signal, e.g., WIMPs in a mass range around 50$-$80 GeV annihilating into $b \bar b$ with the cross section about (1$-$5)$\times 10^{-26}$ cm$^3$/s  \cite{Cuoco:2016eej,Cui:2016ppb}. It happens that the range of DM inferred from the GC gamma-ray excess could coincide with that from antiproton observations, and these two messenger results suggest that the signals may be owing to the same origin from DM annihilations, i.e. WIMPs in a mass range about 50$-$75 GeV mainly annihilating into $b \bar b$ with an annihilation cross section $\sim$ (1$-$3)$\times 10^{-26}$ cm$^3$/s (this is corresponding to about 2$\sigma$ region for the two joint fitting results of Ref. \cite{Cuoco:2017rxb}).

Now, how to realize the DM annihilations suggested above and meanwhile being compatible with constraints from dwarf spheroidal galaxies becomes a crucial question,\footnote{In Ref. \cite{Cline:2017lvv}, the tension becomes relaxed with the assumption that the DM in dwarf spheroidal galaxies may be overestimated, and the similar case was also discussed in Ref. \cite{Cuoco:2017rxb}.} and this is of our concern in this paper. Here we try to relax the tension in mechanism, and consider a scenario that DM pairs annihilate into pairs of unstable particles with the unstable particles mainly decaying into the standard model (SM) $b \bar b$. To satisfy the constraints from dwarf galaxies, a possible solution is DM and the unstable particle being in a multiplet with nearly degenerate masses. The annihilations of DM near the threshold are phase space suppressed today, which is sensitive to the velocity of DM (see e.g. Refs. \cite{Jia:2016pbe,Kopp:2016yji}), and the constraints from dwarf galaxies can be relaxed due to the relatively low velocity. This scheme can be realized via two dark charged pseudoscalar particles $S_d^+$, $S_d^-$ and one neutral pseudoscalar particle $S_d^0$ being in a triplet like SM pions in the hidden sector (see e.g. Ref. \cite{Kopp:2016yji} for more). The DM candidate particles $S_d^+$, $S_d^-$ are stable due to the dark charge, and the unstable neutral particle $S_d^0$ is considered to couple with SM fermions with the couplings proportional to the fermions' masses. A new scalar field is introduced, which mediates the main annihilation process of DM $S_d^+ S_d^-  \rightarrow S_d^0 S_d^0$. As the annihilation of DM today is suppressed, to obtain the indicated DM annihilation cross section, we consider the case of DM annihilating near the resonance. This scheme is also helpful to evade the present stringent constraints from DM direct detections \cite{Akerib:2015rjg,Aprile:2015uzo,Tan:2016zwf,Akerib:2016vxi}.

For thermally freeze-out DM, the DM and SM particles were in the thermal equilibrium for some time in the early Universe, and this can be easily realized for the case of DM directly annihilating into SM particles. For the DM annihilations of concern, the thermal equilibrium is obtained via a small mixing between the new scalar mediator and the SM Higgs boson. This sets an lower bound on the mixing, and the corresponding constraint will be derived. The possible traces from the hidden sector will be discussed, i.e. the production of the new scalar at collider, indirect detections of the $\gamma$ ray line and the search of DM in direct detections.

This work is organized as follows. After this introduction, we briefly give the interactions of a new scalar mediated WIMPs in Section II. Next we will discuss the indicated WIMP annihilations in Section III. Then we give a numerical analysis about traces of the new sector with corresponding constraints in Section IV. The last section is a brief conclusion and discussion.

\section{Interactions of new scalar mediated WIMPs}

In this paper, we consider that two dark charged pseudoscalar particles $S_d^+$, $S_d^-$ and one neutral pseudoscalar particle $S_d^0$ are in a dark triplet, with the stable particles $S_d^+$, $S_d^-$ being DM candidates and the unstable particle $S_d^0$ decaying into SM fermions. This can be obtained under a gauge symmetry and/or a global symmetry, e.g. dark pions in the hidden sector SU($N$) symmetry \cite{Bhattacharya:2013kma,Kopp:2016yji}. A new scalar field $\Phi$ is introduced which couples to the dark triplet particles, while the SM Higgs field $H$ does't directly interact with the dark triplet particles. The effective interactions between $\Phi$ and $S_d^+$, $S_d^-$, $S_d^0$, $H$ are taken as
\begin{eqnarray}
\mathcal {L}_{\Phi}^{\,i} &=&  - \frac{1}{2} \lambda \Phi^2 S_d^+ S_d^-   - \frac{1}{4} \lambda_0 \Phi^2 S_d^0 S_d^0 - \mu \Phi S_d^+ S_d^-   - \frac{1}{2} \mu_0 \Phi S_d^0 S_d^0     \nonumber \\
 && - \lambda_h \Phi^2 ( H^\dag H - \frac{v^2}{2} ) - \mu_h^{} \Phi ( H^\dag H - \frac{v^2}{2} )   \,, \label{fermion-DM}
\end{eqnarray}
where $v \approx$ 246 GeV is the vacuum expectation value, and $\Phi$ is taken to be no vacuum expectation gained \cite{Pospelov:2007mp,Batell:2012mj}. The mass splitting $\Delta$ between $S_d^+$ and $S_d^0$ is $\Delta = m_{S_d^+} - m_{S_d^0}$, and this splitting can be very small due to the symmetry in the dark sector, i.e. $| \Delta |  \ll m_{S_d^+}, m_{S_d^0}$. Here we take $\lambda_0 = \lambda$, $\mu_0 = \mu$ for simplicity. The neutral pseudoscalar $S_d^0$ couples to SM fermions, and the effective form is taken as
\begin{eqnarray}
\mathcal {L}_{S}^{\,i} = \sum_f i g_f \bar{f} \gamma^5 f  S_d^0   \,,
\end{eqnarray}
with $g_f$ being proportional to the fermion's mass.\footnote{ This effective coupling can be obtained, e.g., via interactions in the technicolor like scheme \cite{Weinberg:1975gm,Susskind:1978ms,Farhi:1980xs,Ryttov:2008xe}.}

The $\Phi$ field mixes with the scalar component $h'$ of the Higgs field after the electroweak symmetry breaking, generating the mass eigenstates $\phi$, $h$ in forms of
\begin{eqnarray}
\left (
\begin{array}{c}
  \phi \\
   h
\end{array}
\right ) = \left [
\begin{array}{cc}
   \cos\theta   & \sin\theta  \\
   -\sin\theta  & \cos\theta
     \end{array} \right ]
 \left (
\begin{array}{c}
  \Phi \\
   h'
\end{array}
\right ) \, ,
\end{eqnarray}
where $\theta$ is the mixing angle, with the value
\begin{eqnarray}
\tan 2\theta = \frac{2 v \mu_h}{ m_\Phi^2 - m_{h'}^2 }  \,  \, .
\end{eqnarray}
For the alteration of the Higgs sector being as small as possible, here we consider the case of $\lambda_h \ll 1$ and $| v \mu_h | \ll \min(m_{h'}^2, m_\Phi^2)$. Thus, one has $m_h^{} \simeq m_{h'}^{}$, $m_\phi^{} \simeq m_{\Phi}^{}$, and the mixing angle $\theta$ can be very small compared with unity, i.e. $|\sin \theta| \simeq | \theta| \ll 1$. In addition, the mixing angle can play a crucial role in the equilibrium between the DM sector and SM sector in the early Universe, and this will be discussed in the following.

\section{WIMP annihilations}

Here the WIMP pair $S_d^+ S_d^-$ mainly annihilates into $S_d^0 S_d^0$ with the transition mediated by $\phi$, and $S_d^0$ decays into SM massive fermions. The corresponding annihilation cross section in one particle rest frame is
\begin{eqnarray}
\sigma_{ann} v_r \simeq  \frac{1}{2} \frac{ \beta_f  }{32 \pi (s - 2 m_{S_d^+}^2)} \frac{\mu^4 }{(s -  m_\phi^2)^2 +  m_\phi^2  \Gamma_\phi^2}   \,, \label{dm-ann}
\end{eqnarray}
where $v_r$ is the relative velocity between a WIMP pair, and the factor $\frac{1}{2}$ is due to the $S_d^+ S_d^-$ pair required in annihilations. $s$ is the total invariant mass squared, and $\Gamma_\phi^{}$ is the decay width of $\phi$. $\beta_f$ is a kinematic factor, with $ \beta_f = \sqrt{1- 4 m_{S_d^0}^2 / s }$. The annihilation of a WIMP pair near the threshold is deeply phase space suppressed today. To meet the annihilation cross section indicated by the GC gamma-ray excess and antiproton observations and meanwhile being compatible with the DM relic density (see Appendix \ref{appendix:freeze-out} for the relic density calculations), we consider the WIMPs annihilating near the resonance. In the non-relativistic case, we have $s \simeq 4  m_{S_d^+}^2 +  m_{S_d^+}^2 v_r^2$. The typical $v_r$ of DM today in the Milky Way is $v_r / c \sim 10^{-3}$, and the value is $v_r / c \lesssim 10^{-4}$ in the dwarf galaxies. For WIMP annihilations today, the factor $\beta_f$ can be approximately written as
\begin{eqnarray}
\beta_f \approx \sqrt{2 \frac{\Delta }{m_{S_d^+}} + \frac{1}{4}v_r^2}   \,.
\end{eqnarray}
To be compatible with the constraints from dwarf galaxies, the constraint of the mass splitting is $|\Delta| \ll 10^{-7} m_{S_d^+}$, and thus the factor $\beta_f$ is $\beta_f \approx |v_r| / 2$ (see also Ref. \cite{Kopp:2016yji}).

\begin{figure}[htbp]
\includegraphics[width=0.44\textwidth]{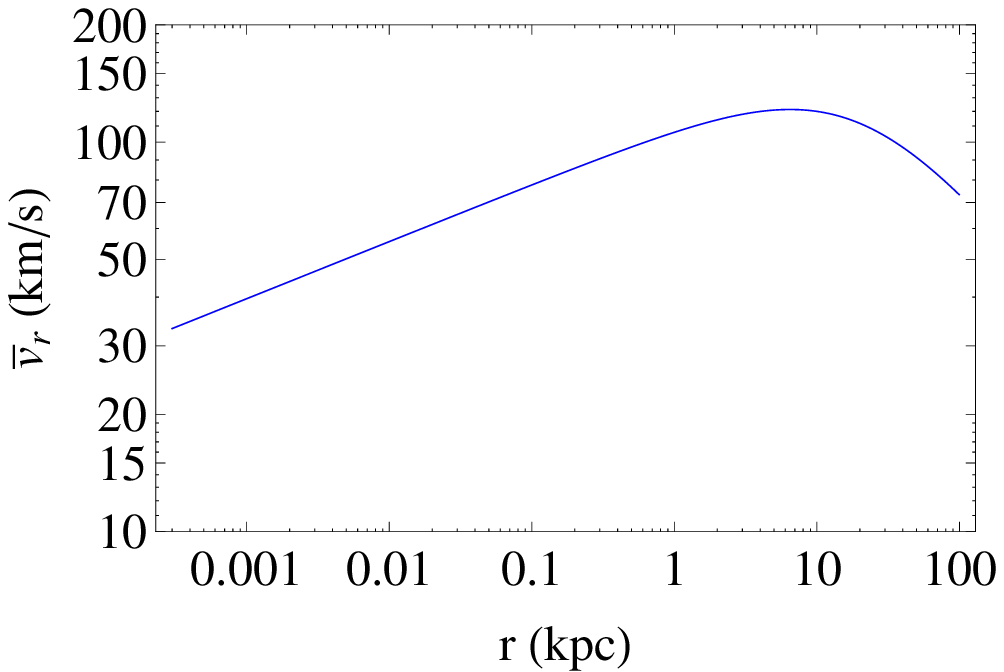} \vspace*{-1ex}  \quad
\includegraphics[width=0.44\textwidth]{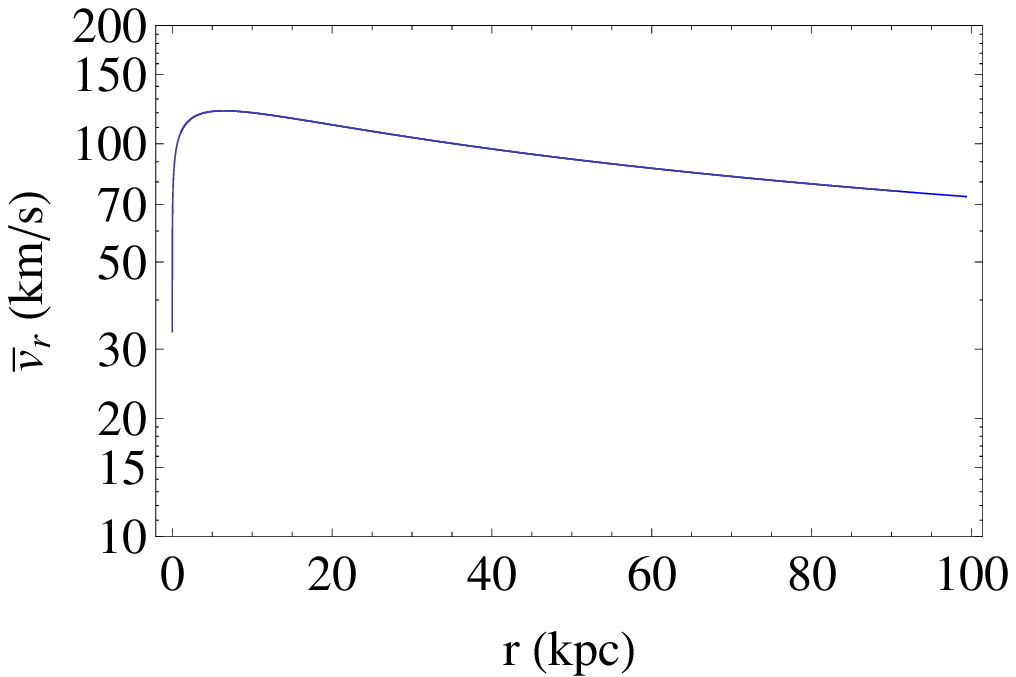} \vspace*{-1ex}
\caption{The result of $\bar{v}_r$ as a function of $r$ in the Milky Way.}\label{vr-r}
\end{figure}

The WIMP pair annihilations today depend on the relative velocity $v_r$. In the Milky Way, the averaged relative velocity $\bar{v}_r$ as a function of $r$ is shown in Fig. \ref{vr-r}. For more details, see Appendix \ref{appendix:vr-today}. For a given position, the averaged annihilation cross section $\langle \sigma_{ann} v_r \rangle_0$ today can be obtained via Eq. (\ref{dm-ann}), with $\beta_f$ replaced by $\bar{\beta}_f$, and $\bar{\beta}_f \simeq |\bar{v}_r| / 2$.

Now we turn to the GC gamma-ray excess via DM annihilations. In the Milky Way, for a solid angle $\Delta \Omega$, the differential flux of the photon $d \Phi_{\gamma} / d E_{\gamma}$ from DM annihilations can be written as
\begin{eqnarray}
\frac{d \Phi_{\gamma}}{d E_{\gamma}} = \int_{\Delta \Omega} \int_{\mathrm{l.o.s.}} \frac{d l d \Omega'}{4 \pi} \frac{\rho_{DM}^2 \langle \sigma_{ann} v_r \rangle_0}{2 m_{DM}^2} \frac{d N_{\gamma}}{d E_{\gamma}} ~. \label{dm-flux}
\end{eqnarray}
When $\langle \sigma_{ann} v_r \rangle_0$ is insensitive to the DM velocity, a $J$-factor can be separated from Eq. (\ref{dm-flux}), with the form
\begin{eqnarray}
J = \int_{\Delta \Omega} \int_{\mathrm{l.o.s.}} \rho_{DM}^2  d l d \Omega'   ~.
\end{eqnarray}
In this paper, due to $\bar{\beta}_f$, the averaged annihilation cross section of DM today is velocity dependent. Here, a weighted $\beta_J^{} \simeq |v_J^{}| / 2$ is introduced, with
\begin{eqnarray}
v_J^{} = \frac{\int_{\Delta \Omega} \int_{\mathrm{l.o.s.}} \bar{v}_r \rho_{DM}^2  d l d \Omega'}{J}   ~.
\end{eqnarray}
Thus, Eq. (\ref{dm-flux}) can be rewritten as
\begin{eqnarray}
\frac{d \Phi_{\gamma}}{d E_{\gamma}} =  \frac{ \langle \sigma_{ann} v_r \rangle_J^{}}{8 \pi m_{DM}^2} \frac{d N_{\gamma}}{d E_{\gamma}} J ~,    \label{gamma-ray}
\end{eqnarray}
where $\langle \sigma_{ann} v_r \rangle_J^{}$ is obtained via Eq. (\ref{dm-ann}), with $\beta_f$ replaced by $\beta_J^{}$.

According to Eq. (\ref{gamma-ray}), to obtain the indicated DM annihilations, an alternative scheme is the WIMP pair annihilations via the process $S_d^+ S_d^-  \rightarrow S_d^0 S_d^0 \rightarrow (b \bar b) (b \bar b)$. Here the WIMP mass is doubled compared with that in the Introduction, and the required annihilation cross section $\langle \sigma_{ann} v_r \rangle_J^{}$ is multiplied by two, i.e. in a range about
\begin{eqnarray}
m_{S_d^+} &\sim& 100-150 ~ \mathrm{GeV}, \nonumber  \\
\langle \sigma_{ann} v_r \rangle_J^{} &\sim& ( 2-6)\times 10^{-26} \mathrm{cm}^3/s ~ .  \nonumber
\end{eqnarray}
This type DM annihilation is of our concern, and it is nearly equivalent to the indicated DM annihilations in the Introduction in kinematics.

Suppose that the distribution of the DM halo is spherical around the GC. For the gamma rays within a 5$^\circ$ cone towards the GC, the main contribution is from the region where the radius $r \lesssim$ 0.74 kpc around the GC. Furthermore, it can be obtained that the distribution of Eq. (\ref{dm-distribution}) is valid for $r \geq 3 \times 10^{-4}$ kpc. The region of interest with $3 \times 10^{-4}$ kpc $ \leq r \lesssim$ 0.74 kpc around the GC can give the main contribution to the 5$^\circ$ cone from the GC. Substituting the corresponding values, the value of $v_J^{}$ is obtained, with
\begin{eqnarray}
v_J^{} \approx 83 ~ \mathrm{km/s }     ~.     \label{vrj}
\end{eqnarray}
Meanwhile, the typical relative velocities in dwarf spheroidal galaxies are less than 15 km/s \cite{Walker:2007ju,Zhao:2016xie,Choquette:2016xsw}. This means that the DM annihilation cross sections in dwarf spheroidal galaxies are just about (or less than) 1/5 of the $\langle \sigma_{ann} v_r \rangle_J^{}$ value at the GC, i.e. the equivalent annihilation cross section (DM directly annihilating into $b \bar{b}$) $\lesssim ( 0.2-0.6)\times 10^{-26}$ cm$^3/$s. Thus, the constraints from dwarf spheroidal galaxies (Refs. \cite{Ackermann:2015zua,Li:2015kag,Fermi-LAT:2016uux}) are relaxed and compatible with the GC gamma-ray excess.

In addition, the indicated DM mass and annihilation cross section in the Introduction can give a joint fit to both the GC gamma-ray excess and the AMS-02 antiproton observations with the assumption of the same annihilation cross section (velocity independent). Here the annihilation cross section of DM is linearly dependent on the relative velocity. It can be seen from Fig. \ref{vr-r} that the typical $\bar{v}_r$ in the Milky Way is about 100 km/s, and thus the parameter space of the GC gamma-ray excess is consistent with the AMS-02 antiproton observations. Moreover, as a rough estimate, a slightly larger typical DM annihilation cross section at significant regions (near the Sun) is also favored by the AMS-02 antiproton observations \cite{Cuoco:2016eej,Cui:2016ppb}, and this suggests that the velocity linearly dependent DM annihilations could give a sightly better fit compared with the velocity independent case. Further explorations are needed about this type DM annihilations.

Considering the annihilations of DM near the resonance, we note $\xi$ = $m_\phi^{} / 2 m_{S_d^+}$, and thus $\xi$ is around 1. The form of Eq. (\ref{gamma-ray}) is feasible in the case of the relative velocity being negligible compared with the mass difference between $m_\phi^{}$ and $2 m_{S_d^+}$, i.e. $(v_J^{} / c )^2 \ll 8 |\xi - 1|$. According to Eq. (\ref{vrj}), this is valid when
\begin{eqnarray}
|\xi - 1| \gtrsim 10^{-7} ~ .   \label{xi-value}
\end{eqnarray}
In addition, the $\sin^2 \theta$ should be small enough to keep the process $S_d^+ S_d^-  \rightarrow S_d^0 S_d^0$ dominant in DM annihilations today. The main SM modes in DM annihilations are $W^+ W^-$, $Z^0 Z^0$ ($h h$ may also be allowed), and these modes are away from the threshold. Note $\mu$ = $k m_{S_d^+}$. For the range of concern, considering Eqs. (\ref{dm-ann}), (\ref{gamma-ray}), (\ref{vrj}), one has that the $\sin^2 \theta$ should be approximately small than the value $5 \times 10^{3} k^2 / m_{S_d^+}^{4} (\mathrm{GeV})$, with $m_{S_d^+}$ in units of GeV. Here, the value
\begin{eqnarray}
\sin^2 \theta \lesssim 10^{-5} k^2 (\frac{100}{m_{S_d^+}} )^4  \label{theta-up}
\end{eqnarray}
is adopted. When $\xi > 1$, to obtain the DM annihilations of concern, the required $\xi - 1$ is of order $10^{-7}$. For the $|\xi - 1|$ value as large as possible within the parameter space, we focus on the case of $\xi < 1$ in the following. Moreover, for the case that the mass of the mediator $m_\phi$ is about twice of the DM mass, this behavior may be naturally realized in other contexts. For example, for the Kaluza-Klein (KK) particles in the universal extra dimension, the masses of the second level KK particles can be twice of the first level KK particles. Thus, pairs of the first level KK particles can annihilate closely to the resonance via the second level KK particles. For more discussions, see e.g. Refs. \cite{Kakizaki:2005en,Ibe:2008ye}.

\section{Numerical analysis with constraints}

Here we give a numerical analysis about the new sector beyond the SM. Some parameters are inputted as follows: $m_t$ = 173.21 GeV, $m_b$ = 4.18 GeV, $m_W$ = 80.385 GeV, $m_Z$ = 91.1876 GeV, $G_F$ = 1.1663787$\times 10^{-5}$ GeV$^{-2}$ \cite{Olive:2016xmw}, and $m_h$ = 125.09 GeV \cite{Aad:2015zhl}.

\subsection{The results with the DM relic density}

\begin{figure}[htbp]
\includegraphics[width=0.44\textwidth]{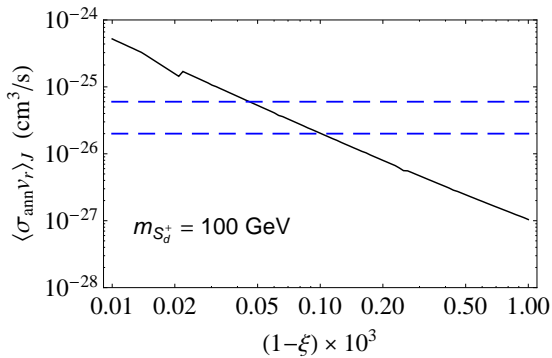} \vspace*{-1ex}
\caption{The relation between $\langle \sigma_{ann} v_r \rangle_J^{}$ and $1 - \xi$ for $m_{S_d^+}$ = 100 GeV. The value of $1 - \xi$ varies in a range of $10^{-5} - 10^{-3}$, and the solid curve is the corresponding value of $\langle \sigma_{ann} v_r \rangle_J^{}$ for a given $1 - \xi$. The upper dashed curve, the lower dashed curve are for the values of $\langle \sigma_{ann} v_r \rangle_J^{}$ = 6$\times 10^{-26}$ cm$^3$/s, 2$\times 10^{-26}$ cm$^3$/s respectively.} \label{sv-today}
\end{figure}

\begin{figure}[htbp]
\includegraphics[width=0.44\textwidth]{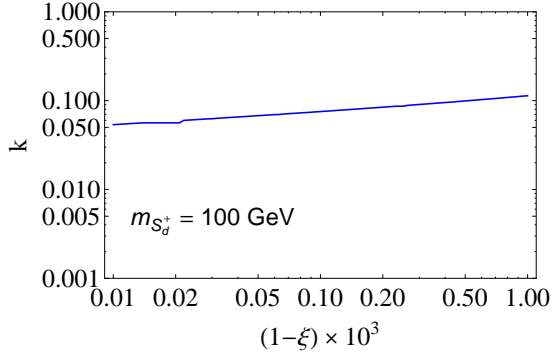} \vspace*{-1ex}
\caption{The relation between $k$ and $1 - \xi$ for $m_{S_d^+}$ = 100 GeV. The value of $1 - \xi$ varies in a range of $10^{-5} - 10^{-3}$, and the solid curve is the corresponding value of $k$ for a given $1 - \xi$.} \label{k-100}
\end{figure}

Here the values of the parameters $\mu$, $\xi$ will be evaluated with the constraint of the DM relic density. The DM relic density today is 0.1197$\pm$0.0042 \cite{Ade:2015xua}. The DM annihilation cross section $\langle \sigma_{ann} v_r \rangle_J^{}$ is sensitive to the value $1 - \xi$ (here $\xi < 1$, as discussed above). For $\xi$ in the range of Eq. (\ref{xi-value}), the decay width $\Gamma_\phi$ can be neglected in DM annihilations when Eq. (\ref{theta-up}) is satisfied. Taking the constraint from the DM relic density, the values of $\langle \sigma_{ann} v_r \rangle_J^{}$, the coupling parameter $k$ as a function of $1 - \xi$ are shown in Fig. \ref{sv-today}, Fig. \ref{k-100} respectively, with $m_{S_d^+}$ = 100 GeV, and $1 - \xi$ varying from $10^{-5}$ to $10^{-3}$. It can be seen that, when $1 - \xi$ changes in a range about (0.45$-$1.0)$\times 10^{-4}$, the $\langle \sigma_{ann} v_r \rangle_J^{}$ varies from about 6$\times 10^{-26}$ cm$^3$/s to 2$\times 10^{-26}$ cm$^3$/s.

\begin{figure}[htbp]
\includegraphics[width=0.44\textwidth]{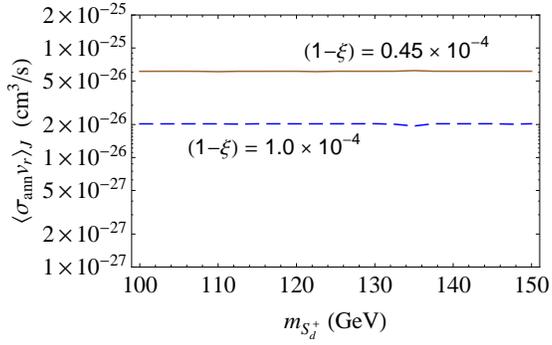} \vspace*{-1ex}
\caption{The values of $\langle \sigma_{ann} v_r \rangle_J^{}$ for two given values of $1 - \xi$. The solid curve, dashed curve are corresponding to $(1 - \xi)$ = 0.45 $\times 10^{-4}$, 1.0 $\times 10^{-4}$ respectively. $m_{S_d^+}$ varies in a range of 100$-$150 GeV.} \label{sigv-m}
\end{figure}

\begin{figure}[htbp]
\includegraphics[width=0.44\textwidth]{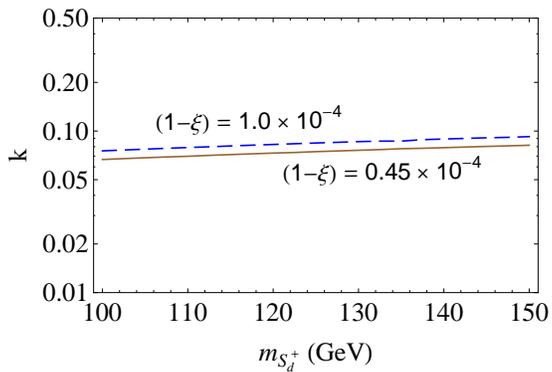} \vspace*{-1ex}
\caption{The values of $k$ for two given values of $1 - \xi$. The solid curve, dashed curve are corresponding to $(1 - \xi)$ = 0.45 $\times 10^{-4}$, 1.0 $\times 10^{-4}$ respectively. $m_{S_d^+}$ varies in a range of 100$-$150 GeV.} \label{k-m}
\end{figure}

For a given $1 - \xi$, the value of $\langle \sigma_{ann} v_r \rangle_J^{}$ is not sensitive to the mass $m_{S_d^+}$ in the WIMP mass range of concern. The result is shown in Fig. \ref{sigv-m} with (1$ - \xi$) = 0.45$ \times 10^{-4}$, 1.0$\times 10^{-4}$, and the corresponding coupling parameter $k$ is shown in Fig. \ref{k-m}. For $m_{S_d^+}$ in a range of 100$-$150 GeV, when $1 - \xi$ changes in a range about (0.45$-$1.0)$\times 10^{-4}$, the $\langle \sigma_{ann} v_r \rangle_J^{}$ varies from about 6$\times 10^{-26}$ cm$^3$/s to 2$\times 10^{-26}$ cm$^3$/s. This parameter range can give an explanation about the GC gamma-ray excess and the AMS-02 antiproton observations.

\subsection{The $\phi$}

\subsubsection{The decay of $\phi$ and constraints from collider}

Due to a small mixing of $\phi$ with the SM Higgs boson, the couplings of $\phi$ with SM particles are Yukawa type interactions. As discussed above, the mass of $\phi$ is about twice of the DM mass, i.e. $m_\phi \sim$ 200$-$300 GeV, and the mass $m_\phi$ being slightly below $2 m_{S_d^+}$ is of our concern. In this case, the main decay products of $\phi$ are $W^+ W^-$, $Z^0 Z^0$. The decay width of $\phi \rightarrow V V $ ($V = W, Z$) is
\begin{eqnarray}
\Gamma_{\phi \rightarrow V V} \simeq \frac{G_F m_\phi^3 \sin^2 \theta}{8 \sqrt{2} \pi \delta_V^{}}   \sqrt{1 - 4 y} (1 - 4 y + 12 y^2)      ~ ,
\end{eqnarray}
with $y = m_V^2 / m_\phi^2$, and $\delta_W^{}$ = 1, $\delta_Z^{}$ = 2. If $m_\phi >$ 250 GeV, the channel $\phi \rightarrow h h$ is also allowed, and the decay width is
\begin{eqnarray}
\Gamma_{\phi \rightarrow h h} \simeq \frac{G_F m_\phi^3 \sin^2 \theta}{16 \sqrt{2} \pi }   \sqrt{1 - \frac{4 m_h^2}{m_\phi^2}} (1 + \frac{2 m_h^2}{m_\phi^2})^2        ~ .
\end{eqnarray}

\begin{figure}[htbp]
\includegraphics[width=0.44\textwidth]{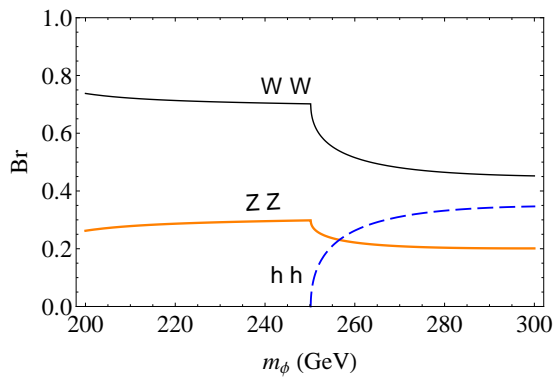} \vspace*{-1ex}
\caption{The branching ratios of $\phi$ decaying into $W^+ W^-$, $Z^0 Z^0$ and $h h$.} \label{branch}
\end{figure}

The branching ratios of the main decay channels of $\phi$ are shown in Fig. \ref{branch}. For $m_{\phi}$ = 200$-$300 GeV, the branching ratio $\mathcal{B}_{\phi \rightarrow Z  Z }$ is about 20\%$-$30\%, and this channel can be tested at LHC. The production of $\phi$ at LHC is mainly through the gluon-gluon fusion, which is very similar to the same mass Higgs boson production, with the corresponding cross section multiplied by $\sin^2 \theta$. For $p p$ collision at the center of mass energy $\sqrt{s}$ = a few TeV, the production cross section of Higgs with masses in a range of 200$-$300 GeV was estimated in Refs. \cite{Baglio:2010ae,Ellis:2017upx}: at $\sqrt{s}$ = 13 TeV, the cross section varies from about 17.5 pb ($m_h$ = 200 GeV) to 9.5 pb ($m_h$ = 300 GeV); at $\sqrt{s}$ = 14 TeV, the cross section varies from about 20.8 pb ($m_h$ = 200 GeV) to 11.2 pb ($m_h$ = 300 GeV). The search of high mass scalar resonance via the decay products of $Z^0 Z^0$ $\rightarrow$ 4$l$ ($l$ = $e$, $\mu$) at $\sqrt{s}$ = 13 TeV was issued by the CMS \cite{CMS:2016ilx} and ATLAS \cite{ATLAS:2016oum} Collaborations, i.e. the observed cross section is a few fb (with the branching ratios included). As a rough estimate, $\sin^2 \theta \lesssim 10^{-2}$ is allowed by the search results. According to Eq. (\ref{theta-up}), the decay products of $\phi$ are buried in the messy background at LHC. Due to the suppression of $\sin^2 \theta$, the search of $\phi$ is challenging in the future TeV scale $e^+ e^-$ collider, as indicated in Ref. \cite{Djouadi:2005gi}.

\subsubsection{The thermal equilibrium constraint}

For thermally freeze-out WIMPs, the WIMPs and SM particles were in the thermal equilibrium in the early Universe, i.e. the reaction rates of WIMPs $\leftrightarrow$ SM particles should be over the expansion rate of the Universe for some time (see. e.g. Refs. \cite{Chu:2011be,Dolan:2014ska,Krnjaic:2015mbs} for more), with
\begin{eqnarray}
\langle \sigma v_r \rangle  n_{\mathrm{eq}} \gtrsim 1.66   \frac{\sqrt{g_{\ast}}  T^2}{m_{\mathrm{Pl}}}      ~, \label{eq-dm-sm}
\end{eqnarray}
where $n_{\mathrm{eq}}$ is the number density in the thermal equilibrium. In the relativistic limit, one has $n_f$ = 3$\zeta (3) g_f T^3 / 4 \pi^2$ for fermions, and $n_b$ = $\zeta (3) g_b T^3 /  \pi^2$ for bosons. Here we consider the case that the reaction rate of SM particles $\rightarrow$ WIMPs can be over the expansion rate of the Universe at some time after the electroweak symmetry breaking. The transitions mainly contributed by the mixing between the new scalar mediator and the Higgs boson. For details about the transitions, see Appendix \ref{SM-DM}. Thus, Eq. (\ref{eq-dm-sm}) sets an lower bound on the mixing angle $\theta$.

\begin{figure}[htbp]
\includegraphics[width=0.44\textwidth]{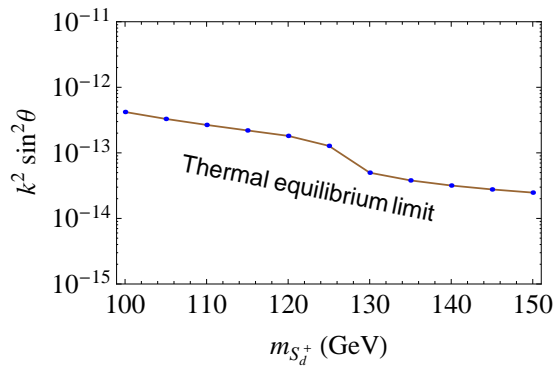} \vspace*{-1ex}
\caption{The lower limit of $k^2 \sin^2 \theta$ from the thermal equilibrium, with $m_{S_d^+}$ varying in a range of 100$-$150 GeV.} \label{t-eq-l}
\end{figure}

For the reaction rate, the annihilation cross section at least decreases as $1/s$ at a high energy, while the number density is exponentially suppressed at a low energy. In this paper $m_\phi \sim 2 m_{S_d^+}$ and $m_{S_d^+} \sim$ 100$-$150 GeV, as the transitions mediated by $\phi$ close to the resonance are significantly enhanced, we take the temperature scale $T \sim m_h$ and just consider the transitions mediated by $\phi$ with the contributions from $W^+ W^-$, $Z^0 Z^0$ and $h h$ pairs in calculations. The lower limit of $k^2 \sin^2 \theta$ is shown in Fig. \ref{t-eq-l}.

\subsection{The $S_d^0$}

Here we assume that the lifetime of $S_d^0$ is less than (or similar to) the time scale from the beginning of the Big Bang to the moment that the temperature of the Universe cooling to $m_{S_d^0}$, and thus $S_d^0$ is in relativistic when its decay occurs. In this case, the number density of $S_d^0$ can be of its equilibrium value during $S_d^+, S_d^-$ freeze out \cite{Kawasaki:1992kg,Kopp:2016yji}. This sets an lower bound on the couplings of $S_d^0$ to SM fermions. In the early Universe of the radiation dominant epoch, the temperature $T$ can be written as a function of time $t$,
\begin{eqnarray}
T = (\frac{16 \pi^3 }{45} G g_\ast)^{- 1/4} ~  t^{-1/2}       ~ ,
\end{eqnarray}
where $G$ (=1/$m_{\mathrm{Pl}}^2$) is the Newton's constant of gravitation. At $T \sim m_{S_d^0}$, the effective lifetime $\tau_{eff}$ of $S_d^0$ is
\begin{eqnarray}
\frac{1}{\tau_{eff}} \simeq \frac{m_{S_d^0}}{\langle E_{S_d^0}\rangle} \sum_f \frac{g_f^2 N_c m_{S_d^0}}{8 \pi} \sqrt{1- 4 m_f^2/m_{S_d^0}^2}        ~ ,
\end{eqnarray}
where the time dilation effect is considered, and $\langle E_{S_d^0}\rangle$ is the averaged energy of $S_d^0$. At $T$ = $m_{S_d^0}$, one has $\langle E_{S_d^0}\rangle$ $\approx$ 3.25 $T$. The main decay product of $S_d^0$ is $b \bar{b}$, and thus we have
\begin{eqnarray}
g_b^2 \gtrsim  7.40 \times 10^{-18}  \sqrt{g_\ast}  m_{S_d^0} (\mathrm{GeV})  ~ .
\end{eqnarray}

\begin{figure}[htbp]
\includegraphics[width=0.44\textwidth]{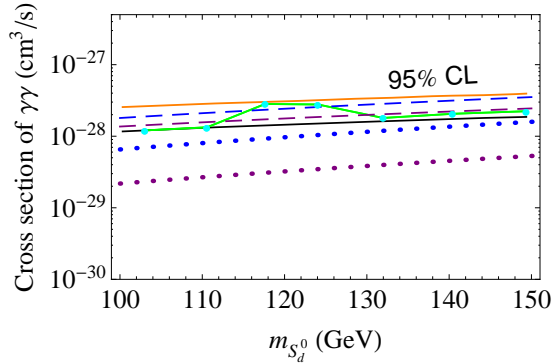} \vspace*{-1ex}
\caption{The annihilation cross section of $\gamma \gamma$ (the value of 2$\langle \sigma_{ann} v_r \rangle_J^{} \times$  $\mathcal{B}_{S_d^0 \rightarrow \gamma \gamma}$), with the energy of the gamma-ray line $\sim m_{S_d^0} /2$ and $m_{S_d^0}$ varying in a range of 100$-$150 GeV. The lower dotted curve, upper dotted curve are for the case of $\langle \sigma_{ann} v_r \rangle_J^{}$ = 2$\times 10^{-26}$ cm$^3$/s, 6$\times 10^{-26}$ cm$^3$/s respectively. The lower solid curve, upper solid curve are the expectation value of the gamma-ray line limit, the upper 95\% containment of the expectation value respectively, for the DM profile of NFWc R3  \cite{Ackermann:2015lka}. The two dashed curves are the expectation value of the gamma-ray line limit + the contribution from DM annihilations, with the lower one, upper one for the case of $\langle \sigma_{ann} v_r \rangle_J^{}$ = 2$\times 10^{-26}$ cm$^3$/s, 6$\times 10^{-26}$ cm$^3$/s respectively. The solid-dotted curve is the observed limit \cite{Ackermann:2015lka}.} \label{s-gamma}
\end{figure}

The $S_d^0$ particle decays into $\gamma \gamma$ via charged fermion loops, and this is constrained by the GC gamma-ray line observation. The decay width is \cite{Djouadi:2005gj}
\begin{eqnarray}
\Gamma_{S_d^0 \rightarrow \gamma \gamma} = \frac{\alpha^2 m_{S_d^0}^3}{256 \pi^3} \mid  \sum_f \frac{N_c Q_f^2 g_f F_{S_d^0}(\tau_f)}{m_f} \mid^2 ~ ,
\end{eqnarray}
where $\tau_f$ = $m_{S_d^0}^2 / 4 m_f^2$, and
\begin{eqnarray}
F_{S_d^0}(\tau_f) = \frac{2}{\tau_f} \times \bigg \{ \begin{array}{cc}
  \arcsin^2 \sqrt{\tau_f} ~ , ~ & \tau_f \leq 1  \\
  - \frac{1}{4} \big [ \log \frac{1 + \sqrt{1 - \tau_f^{-1}}}{1 - \sqrt{1 - \tau_f^{-1}}} - i \pi \big ]^2 , &   \tau_f > 1
\end{array}
~ ~ .
\end{eqnarray}
The top quark gives the main contribution to the decay width. Though the value of $g_f$ is unknown, the branching ratio $\mathcal{B}_{S_d^0 \rightarrow \gamma \gamma}$ is set yet. Here, the annihilation cross section of the $\gamma \gamma$ is about 2$\langle \sigma_{ann} v_r \rangle_J^{} \times$  $\mathcal{B}_{S_d^0 \rightarrow \gamma \gamma}$, with the energy of the gamma-ray line $\sim m_{S_d^0} /2$ (and also $\sim m_{S_d^+} /2$). The GC gamma-ray line was searched by the Fermi-LAT, and within 95\% containment of the expectation value, no significant spectral line was found \cite{Ackermann:2015lka}. This limit can be employed to set an upper limit about the DM annihilation of concern, and the revised upper limit can be obtained via the DM mass doubled and the corresponding upper limit multiplied by four compared with that in Ref. \cite{Ackermann:2015lka}. The result is shown in Fig. \ref{s-gamma}, and it can be seen that the gamma-ray line from DM annihilations is allowed by the present observations within 95\% containment.

In addition, the $S_d^0$-like particle search at collider can give an upper limit about the coupling $g_f$, and here we give a brief discussion about it. For the $S_d^0$-b coupling, the constraint about $g_b$ were discussed in Refs. \cite{Berlin:2015wwa,Fan:2015sza}, with $g_b \lesssim 0.1$, and this constraint is mild. Consider that the couplings of $S_d^0$ with SM fermions are smaller compared with that of the Higgs boson, and a limit $g_b^2 \lesssim 10^{-3}  \sqrt{2} G_F m_b^2$ is taken here. This $S_d^0$-like particle is allowed by the diphoton observation at LHC \cite{Aad:2014eha}.

\subsection{DM direct detection}

Now we turn to the direct detection of WIMPs. The WIMP-nucleon spin-independent elastic scattering mediated by $h$ and $\phi$ is
\begin{eqnarray}
\sigma_{\mathrm{el}} \simeq \frac{\sin^2 \theta k^2 m_{S_d^+}^2 g_{hNN}^2 m_N^2}{4 \pi ( m_{S_d^+} + m_N)^2} ( \frac{1}{ m_h^2} - \frac{1}{m_\phi^2 } )^2 ~ ,
\end{eqnarray}
where $m_N$ is the nucleon mass. $g_{hNN}^{}$ is the effective Higgs-nucleon coupling, and $g_{hNN}^{}$ $ \simeq$ 1.1$\times 10^{-3}$ \cite{Cheng:2012qr} is adopted here.\footnote{See e.g. Refs. \cite{Ellis:2000ds,Gondolo:2004sc,He:2008qm,Alarcon:2011zs,Cline:2013gha} for more.}

\begin{figure}[htbp]
\includegraphics[width=0.44\textwidth]{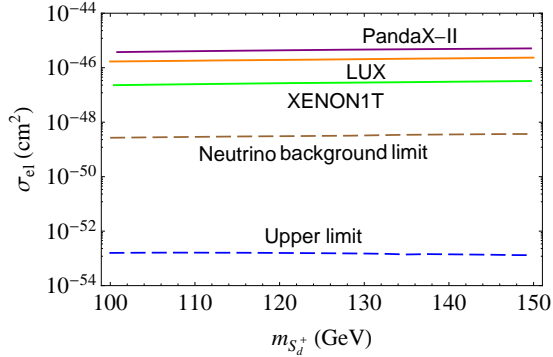} \vspace*{-1ex}
\caption{The result of the elastic scattering cross section $\sigma_{\mathrm{el}}$. The solid curves from top to bottom are the upper limits set by DM direction detections of PandaX-II \cite{Tan:2016zwf}, LUX \cite{Akerib:2016vxi} and XENON1T \cite{Aprile:2015uzo}. The upper dashed curve, lower dashed curve are the lower detection limit set by the neutrino background \cite{Billard:2013qya}, the upper limit of the DM scattering from Eq. (\ref{theta-up}), respectively.} \label{scatt-cs}
\end{figure}

The recent DM direct detections of XENON1T \cite{Aprile:2015uzo}, PandaX-II \cite{Tan:2016zwf} and LUX \cite{Akerib:2016vxi} set stringent constraints on WIMP type DM. To obtain the upper limit of $\sigma_{\mathrm{el}}$, the value of $k$ is taken for the case of (1$-$$\xi$) = $1.0 \times 10^{-4}$ in Fig. \ref{k-m}. Considering the upper limit of Eq. (\ref{theta-up}), the result is shown in Fig. \ref{scatt-cs}.
The expected upper limit of the DM scattering is far below the neutrino background estimated in Ref. \cite{Billard:2013qya}, and the DM of concern escapes the future direct detection experiments.

\section{Conclusion and discussion}

In this paper, the DM annihilation of $S_d^+ S_d^-  \rightarrow S_d^0 S_d^0$ mediated by $\phi$ with $S_d^0$ quickly decaying into $b \bar{b}$ has been studied to explain the GC gamma-ray excess and AMS-02 antiproton observations, with $m_{S_d^+}$ in a range of 100$-$150 GeV and the DM annihilation cross section $\langle \sigma_{ann} v_r \rangle_J^{}$ = 2$-$6 $\times 10^{-26}$ cm$^3$/s. In this scenario, the particles $S_d^+$, $S_d^-$ and $S_d^0$ are in a triplet in hidden sector with degenerate masses. The annihilation cross section of DM today is linearly dependent on the relative velocity $v_r$, and thus constraints from the dwarf spheroidal galaxies are suppressed and relaxed. With the indication of the GC gamma-ray excess, we consider the DM annihilating near the resonance, and the weighted relative velocity $v_J$ $\approx$ 83 km/s is derived.

The values of the coupling parameter $k$, the mass ratio parameter $\xi$ are derived with the constraint of the DM relic density. In the DM mass range of concern, when $1 - \xi$ varies in a range of (0.45$-$1.0) $\times 10^{-4}$, the corresponding $\langle \sigma_{ann} v_r \rangle_J^{}$ varies from about 6 $\times 10^{-26}$ cm$^3$/s to 2 $\times 10^{-26}$ cm$^3$/s. This is favored by the joint results of the GC gamma-ray excess and the AMS-02 antiprotons. In addition, by a rough estimate, it suggests that the velocity linearly dependent DM annihilations could give a sightly better fit compared with the usual assumption of velocity independent. Moreover, the case that the mass of the mediator is about twice of the DM mass, this may be related to other contexts, such as the universal extra dimension \cite{Kakizaki:2005en}, or other undiscovered symmetries. Further explorations are needed for the velocity linearly dependent DM annihilations.

The upper limit of the mixing angle $\theta$ set by LHC is mild, and the search of $\phi$ particle is challenging at future collider experiment. An upper limit on the DM-nucleon elastic scattering cross section is set by the upper limit of $\theta$ from the DM main annihilation process $S_d^+ S_d^- \rightarrow S_d^0 S_d^0$, and this limit is far below the neutrino background in direct detections. The thermal equilibrium in the early Universe sets a lower limit on $k \sin \theta$, which has been calculated. Though traces from the new sector are difficult to be disclosed via the search at collider and the DM direct detection, the indirect search of the gamma-ray line from $S_d^0$'s decay has the potential to shed light on DM annihilations, with the energy of the gamma-ray line $\sim m_{S_d^0} /2$ (about 50$-$75 GeV). We look forward to more precise observations on the GC gamma-ray line, with the results from the Fermi-LAT \cite{Ackermann:2015lka}, the Dark Matter Particle Explorer (DAMPE) \cite{Chang:2014}, the Cherenkov Telescope Array (CTA) \cite{Acharya:2013sxa}, the High Energy cosmic-Radiation Detection (HERD) \cite{Zhang:2014qga,Huang:2015fca} and the GAMMA-400 \cite{Galper:2012fp,Topchiev:2017a}.

\acknowledgments \vspace*{-3ex} This work was supported by National Natural Science Foundation of China  under Contract No. 11505144, and the Research Fund for the Doctoral Program of the Southwest University of Science and Technology under Contract No. 15zx7102.

\appendix

\section{The abundance of DM}
\label{appendix:freeze-out}

The thermally averaged annihilation cross section of DM at temperature $T$ is \cite{Gondolo:1990dk,Cannoni:2013bza}
\begin{eqnarray}
\langle \sigma_{ann} v_r \rangle &=& \frac{2 x}{K_2^2(x)} \int_0^{\infty} \mathrm{d} \varepsilon \sqrt{\varepsilon} (1+2\varepsilon) ~ \nonumber \\
&& \times K_1 (2 x \sqrt{1 + \varepsilon})  \sigma_{ann} v_r  ~ ,
\end{eqnarray}
with $\varepsilon=(s-4m_{S_d^+}^2)/4m_{S_d^+}^2$, and $x=m_{S_d^+} / T$. $K_i$ is the modified Bessel function of order $i$. For thermally freeze-out DM at temperature $T_f$, the thermally averaged annihilation cross section links to the DM relic density $\Omega_D$ today via the relation \cite{Kolb:1990,Griest:1990kh}
\begin{eqnarray}
\Omega_D h^2 \simeq \frac{1.07 \times 10^9 ~ \mathrm{GeV}^{-1}}{J_{ann} \sqrt{g_\ast} m_{\mathrm{Pl}}}  ~ , \label{DM-freeze-out}
\end{eqnarray}
with
\begin{eqnarray}
J_{ann} = \int_{x_f^{}}^{\infty} \frac{\langle \sigma_{ann} v_r \rangle}{x^2}  \mathrm{d}x ~ ,
\end{eqnarray}
and the parameter $x_f^{}$ ($x_f^{} = m_{S_d^+} / T_f^{}$) can be approximately written as
\begin{eqnarray}
x_f^{} \simeq \mathrm{ln} ~ 0.038 \frac{g m_{\mathrm{Pl}} m_{S_d^+} \langle \sigma_{ann} v_r \rangle}{\sqrt{g_\ast ~ x_f^{}} } ~.
\end{eqnarray}
Here $h$ is the reduced Hubble constant (in units of 100 km s$^{-1}$ Mpc$^{-1}$), and $g_\ast$ is the effective number of the relativistic degrees of freedom at temperature $T_f^{}$. $m_{\mathrm{Pl}}$ is the Planck mass with $m_{\mathrm{Pl}}$ = 1.22 $\times 10^{19}$ GeV, and $g$ is the degrees of freedom of DM. The value of $x_f^{}$ is about 20. Here the DM particles $S_d^+, S_d^-$ are non-relativistic when they freeze out. In this paper, we consider the case that $S_d^0$ remains in thermal equilibrium with SM particles before $S_d^+ S_d^-$ freeze out,\footnote{For other case, see e.g. discussions in Ref. \cite{Dror:2016rxc}.} as discussed in Refs. \cite{DAgnolo:2015ujb,Kopp:2016yji}. The abundances of $S_d^+, S_d^-$ and $S_d^0$ are nearly to be the equilibrium abundances before $S_d^+, S_d^-$ freeze out \cite{DAgnolo:2015ujb,Kawasaki:1992kg,Kopp:2016yji}, and Eq. (\ref{DM-freeze-out}) is available.

\section{The relative velocity}
\label{appendix:vr-today}

For a WIMP pair with velocities $v_1$, $v_2$ in the GC, dwarf spheroidal galaxies or the whole Milky Way, the averaged annihilation cross section today is
\begin{eqnarray}
\langle \sigma_{ann} v_r \rangle_0  = \frac{1}{2} \int_0^{v_{esc}} \mathrm{d} v_1 \int_0^{v_{esc}} \mathrm{d} v_2 \int_{-1}^{1} \mathrm{d} \cos \theta f(v_1) f(v_2) \sigma_{ann} v_r ~,
\end{eqnarray}
where the relative velocity $v_r$ can be obtained via the relation $v_r$=$\sqrt{v_1^2 + v_2^2 - 2 v_1 v_2 \cos \theta}$. The escape velocity $v_{esc}$ is radial position $r$ dependent, and one has \cite{Cirelli:2010nh}
\begin{eqnarray}
v_{esc}^2 = 2 \int_r^{\infty} \frac{\mathrm{d}r}{r} v_c^2 (r) ~,
\end{eqnarray}
with $v_c (r)$ being the circular velocity (see Refs. \cite{Sofue:2000jx,Sofue:2013kja} for the value of $v_c$ at a given $r$ in the Milky Way). $f(v)$ is the velocity distribution, and here a Maxwell-Boltzmann distribution is adopted \cite{Robertson:2009bh,Cirelli:2010nh,Choquette:2016xsw}
\begin{eqnarray}
f(v) = \frac{3 \sqrt{6}}{\sqrt{\pi} \sigma^3_v} v^2 e^{-3 v^2 / 2 \sigma^2_v} ~, \label{dm-distribution}
\end{eqnarray}
where $\sigma_v$ is the velocity dispersion at a given $r$. For the Galactic DM, the fitting form is
\begin{eqnarray}
\sigma_v^3 (r) = v_0^3 (\frac{r}{r_s})^{\chi} \frac{\rho(r)}{\rho_0} ~,
\end{eqnarray}
where $\chi$ = 1.64 is adopted with the baryon contributions included \cite{Cirelli:2010nh}, and $v_0$ = 130 km/s is taken \cite{Battaglia:2005rj,Dehnen:2006cm}. A NFW density profile of DM is adopted
\begin{eqnarray}
\rho (r) = \frac{\rho_0}{(\frac{r}{r_s})^{\gamma} (1 + \frac{r}{r_s})^{3 - \gamma}} ~,
\end{eqnarray}
with $r_s$ = 20 kpc, and $\gamma$ = 1.2 being adopted for the best fit result in Refs. \cite{Daylan:2014rsa,Calore:2014xka}. The distance from the Sun to the GC is $R_{\odot} \simeq$ 8.5 kpc, and $\rho_0$ is taken for the DM density near the Sun being $\sim$ 0.4 GeV/cm$^3$. An averaged $\bar{\beta}_f \simeq |\bar{v}_r| / 2$ is introduced, with the averaged relative velocity
\begin{eqnarray}
\bar{v}_r  = \frac{1}{2} \int_0^{v_{esc}} \mathrm{d} v_1 \int_0^{v_{esc}} \mathrm{d} v_2 \int_{-1}^{1} \mathrm{d} \cos \theta f(v_1) f(v_2) v_r ~.
\end{eqnarray}

\section{The transition of SM$\to$DM}
\label{SM-DM}

Consider the $\phi$ mediated transitions of SM particles $\rightarrow$ WIMPs first. For each SM fermion specie, the annihilation cross section is
\begin{eqnarray}
 \sigma v_r (f \bar f) = \frac{\lambda_{SM}^2 \mu^2 (s -  4 m_f^2)}{32 \pi (s - 2 m_f^2) }    \frac{ \sqrt{1- 4 m_{S_d^+}^2 / s } }{(s -  m_\phi^2)^2 +  m_\phi^2  \Gamma_\phi^2}   ~,
\end{eqnarray}
with $\lambda_{SM}$ = $\sin \theta  m_f (\sqrt{2}  G_F)^{1/2}$. For the SM massive vector boson pair $V V$ ($V = W, Z$), the annihilation cross section is
\begin{eqnarray}
 \sigma v_r (V V) &=& \frac{\sin^2 \theta \mu^2 \sqrt{2} G_F  s^2}{144 \pi (s - 2 m_V^2) }    \frac{ \sqrt{1- 4 m_{S_d^+}^2 / s } }{ (s -  m_\phi^2)^2 +  m_\phi^2  \Gamma_\phi^2}  \nonumber \\
 &&   \times (1 - \frac{4 m_V^2}{s} + \frac{12 m_V^4}{s^2})    ~.
\end{eqnarray}
For the $h h$ pair, the annihilation cross section is
\begin{eqnarray}
 \sigma v_r (h h) &=& \frac{\sin^2 \theta \mu^2 \sqrt{2} G_F m_\phi^4}{16 \pi (s - 2 m_h^2) }    \frac{ \sqrt{1- 4 m_{S_d^+}^2 / s } }{ (s -  m_\phi^2)^2 +  m_\phi^2  \Gamma_\phi^2}  \nonumber \\
 &&   \times (1 + \frac{2 m_h^2}{m_\phi^2})^2    ~.
\end{eqnarray}
The transitions mediated by $h$ are similar to that of $\phi$, just with $m_\phi$, $\Gamma_\phi$ replaced by $m_h$, $\Gamma_h$, respectively. For a 125 GeV SM Higgs boson, the total width is $\Gamma_h$ = 4.07 $\times 10^{-3}$ GeV \cite{Olive:2016xmw,Denner:2011mq}.

\end{document}